\documentclass[letters,useAMS,usenatbib]{mn2e}
\usepackage{graphicx}

\def\lesssim{\mathrel{\hbox{\rlap{\hbox{\lower4pt\hbox{$\sim$}}}\hbox{$<$}}}}
\def\gtrsim{\mathrel{\hbox{\rlap{\hbox{\lower4pt\hbox{$\sim$}}}\hbox{$>$}}}}

\def\gtrsim{\mathrel{\hbox{\rlap{\hbox{\lower4pt\hbox{$\sim$}}}\hbox{$>$}}}}

\usepackage{lscape}

\begin{document}

\title[A possibly old galaxy at $z=6.027$ lensed by Abell 383]{
Discovery of a possibly old galaxy at $z=6.027$, multiply imaged by the massive cluster Abell 383
}
\author[Richard et al.]{
 \parbox[h]{\textwidth}{
Johan Richard$^{1,2}$\thanks{E-mail:
johan.richard@univ-lyon1.fr}, Jean-Paul Kneib$^{3}$, 
Harald Ebeling$^{4}$, Daniel P. Stark$^{5}$, Eiichi Egami$^{6}$, Andrew K. Fiedler$^{6}$}
\vspace{6pt}\\
$^{1}$CRAL, Observatoire de Lyon, Universit\'e Lyon 1, 9 Avenue Ch. Andr\'e, 69561 Saint Genis Laval Cedex, France\\
$^{2}$Dark Cosmology Centre, Niels Bohr Institute, University of
Copenhagen, Juliane Maries Vej 30, DK-2100 Copenhagen, Denmark\\
$^{3}$Laboratoire d'Astrophysique de Marseille, CNRS- Universit\'e Aix-Marseille, 38 rue F. Joliot-Curie, 13388 Marseille Cedex 13, France\\
$^{4}$Institute of Astronomy, University of Hawaii, Honolulu, HI 96822, USA\\
$^{5}$Institute of Astronomy, University of Cambridge, Madingley Road, Cambridge CB3 0HA, UK\\
$^{6}$Steward Observatory, University of Arizona, 933 N. Cherry Avenue, Tucson, AZ 85721, USA\\
}

\date{Accepted . Received ; in original form }

\pagerange{\pageref{firstpage}--\pageref{lastpage}} \pubyear{2011}

\maketitle

\label{firstpage}

\begin{abstract}

We report the discovery of a unique $z=6.027$ galaxy, multiply imaged by
the cluster Abell 383 and detected in new \textit{Hubble Space Telescope}
ACS and WFC3 imaging, as well as in {\it Warm Spitzer} observations.  This galaxy
was selected as a pair of $i$-dropouts; its suspected high redshift was confirmed by
the measurement of a strong Lyman-$\alpha$ line in both images using
Keck/DEIMOS.  Combining {\it Hubble} and {\it Spitzer} photometry
after correcting for contamination by line emission (estimated to be a small effect), we identify a strong Balmer break
of 1.5 magnitudes. 
Taking into account the magnification factor of
$11.4\pm1.9$ (2.65$\pm$0.17 mag) for the brightest image, the
unlensed $AB$ magnitude for the source is 27.2$\pm$0.05 in the H band, corresponding to a 0.4 L$^{*}$ galaxy, and 25.7$\pm$0.08 at 3.6 $\mu$m.
The UV slope is consistent with $\beta\sim2.0$, and from the
rest-frame UV continuum we measure a current star formation rate of
2.4$\pm1.1$ M$_{\odot}$/yr. The unlensed half-light radius is measured
to be 300 pc, from which we deduce a star-forming surface density of 
$\sim$10 M$_{\odot}$ yr$^{-1}$ kpc$^{-2}$. The Lyman-$\alpha$
emission is found to be extended over $\sim$3\arcsec along the slit, corresponding
to $\sim$5 kpc in the source plane. This can be explained by the
presence of a much larger envelope of neutral hydrogen around the
star-forming region. Finally, fitting the spectral energy distribution
using 7 photometric data points with simple SED models, we derive the 
following properties: very little reddening, an inferred stellar mass of $M^*$=6\ 10$^{9}$
M$_\odot$, and an inferred age of $\sim$800 Myrs (corresponding to a redshift of
formation of $\sim18$). The star-formation rate
of this object was likely much stronger in the past than at the time of observation, 
suggesting that we may be missing a fraction of galaxies at z$\sim$6 which 
have already faded in rest-frame UV wavelengths.

\end{abstract}


\section{Introduction}

One of the most important topics of modern cosmology is the nature
of the sources that reionized the intergalactic medium in the
early Universe. A lot of progress has been made in
characterizing the bright end of the population of high-redshift galaxies through
the selection of \textit{optical dropouts} using both space- and
ground-based data (e.g. \citealt{Bouwens06,McLure}).  In particular,
the luminosity function of galaxies at $z\sim6$ is well constrained down
to 0.1 L$^*$ thanks to very deep observations of fields such as GOODS and the
Hubble Ultra Deep Field (UDF), which provided over $500$ $i$-dropouts
\citep{Bouwens06,Bouwens07}.

Solely based on photometric selection, the observed trend implies
highly significant evolution at the bright end of the
luminosity function ($>~L^*$) between $z\sim4$ and $z\sim7$,
with a large decrease in the number of luminous sources. If this
trend extends to higher redshifts, there might not be a sufficient
number of ionizing photons to explain the observed change of the
intergalactic medium.  One possible solution to this problem would be the presence
of a high number of low-mass/low-luminosity sources at
these redshifts, which would contribute to the bulk of the ionizing
radiation. However, only a small fraction of the currently known dropouts at $z\sim6$
have luminosities of 0.1-0.5$L^*$ (L$^*$ corresponds to $\sim$26.3 AB at
$z\sim6$), and even fewer are spectroscopically confirmed
\citep{Stark10,Stark11}.

The newly installed Wide Field Camera 3 (WFC3) onboard the \textit{Hubble
Space Telescope} (HST) enables deep near-infrared imaging that is very fast
compared to any other instrument. So far most of the high-redshift work 
using WFC3 has been performed on deep fields, the
UDF in particular, and numerous and convincing candidates for galaxies at redshifts 
$z\sim7$ \citep{McLure,Bunker2010,Finkelstein10} or even $z\sim10$ \citep{Bouwens11} 
have been identified. However, the typical
magnitudes of these objects ($\sim$27-28, and $\sim29$ for the $z\sim10$ candidate) 
make ground-based spectroscopy challenging. 
Combining WFC3 and the \textit{Spitzer Space Telescope} (SST) at longer wavelengths allows us to study the
star-formation history of these high-redshift galaxies, and to derive
constraints on the age of the stellar population and the total stellar mass, thanks to the high
sensitivity of the first two IRAC channels \citep[e.g.,][]{Egami}.  In
particular, \citet{Stark09} measured the masses of optical dropouts
from $z\sim4$ and $z\sim6$, and found no significant evolution in the
ratios of stellar mass to UV light. However, ages and stellar masses are
best constrained in the brightest IRAC detections and quite uncertain for faint objects.

A complementary and very powerful approach is to observe the high-redshift
Universe through massive cluster lenses. Gravitational lensing
offers the opportunity to perform high signal-to-noise observations of
galaxies that would otherwise be too faint and distant for quantitative study.  Although the intrinsic lensed volume probed by any individual cluster observation is small compared to its unlensed equivalent, 
gravitational amplification by clusters likely represents a unique way to perform ground-based spectroscopy of objects
that are intrinsically as faint as mag$\sim$30 \citep{Ellis01,Kneib04,Richard08}.

Combining the resolution and sensitivity of the Wide-Field Camera 3 (WFC3) and the
Advanced Camera for Surveys (ACS) 
with the magnification afforded by clusters used as \textit{gravitational telescopes} the 
recently approved Multi-Cycle Treasury ``CLASH" cluster program on
HST (PI: Postman; 25 clusters; 524 orbits)  offers a fresh opportunity to find 
and study high-redshift galaxies. Two thirds of the CLASH clusters have also been 
observed with SST/IRAC as part of our SST Warm Mission Exploratory
Science program, ``The IRAC Lensing Survey'' (PI: Egami; 53 clusters;
593 hours), which obtains deep (5 hours/band) images in the two
shorter-wavelength IRAC channels (3.6 and 4.5$\mu$m).  Together, these
two large HST and SST observing programs facilitate a detailed study of
the physical properties of distant lensed galaxies.

In this letter, we report the discovery of a high-redshift $i$-dropout
in Abell 383 (ID 12065), which we confirmed to be at $z=6.027$ based on Keck
spectroscopy. We summarize the observations performed in Section 2
and present results of our photometric and spectroscopic analysis
in Section 3. Conclusions are given in Section 4. Magnitudes are
quoted in the AB system. At $z=6$, 1\arcsec\ on the sky corresponds to a
projected distance of 5.710 kpc in the adopted cosmology
($\Omega_m=0.3$, $\Omega_\Lambda=0.7$, $h=0.7$).

\section{Imaging and Spectroscopy}

\begin{table*}
\begin{tabular}{lllllllll}
Image & F775W & F814W & F850LP & F110W & F125W & F160W & 3.6$\mu$m & 4.5$\mu$m \\
\hline
5.1 & $>26.48$ & 25.63$\pm$0.14 & 24.54$\pm$0.12 & 24.65$\pm$0.05 & 24.57$\pm$0.08 & 24.66$\pm$0.05 & 23.10$\pm$0.08 & 22.91$\pm$0.08 \\
5.2 & $>26.48$ & 26.41$\pm$0.23 & 25.35$\pm$0.21 & 25.16$\pm$0.08 & 25.18$\pm$0.08 & 25.15$\pm$0.07 & N/A            & N/A    \\
\end{tabular}
\caption{\label{photometry}Summary of the measured photometry for both images.}
\end{table*}

\begin{figure*}
\begin{minipage}{8cm}
\includegraphics[width=8cm,angle=0]{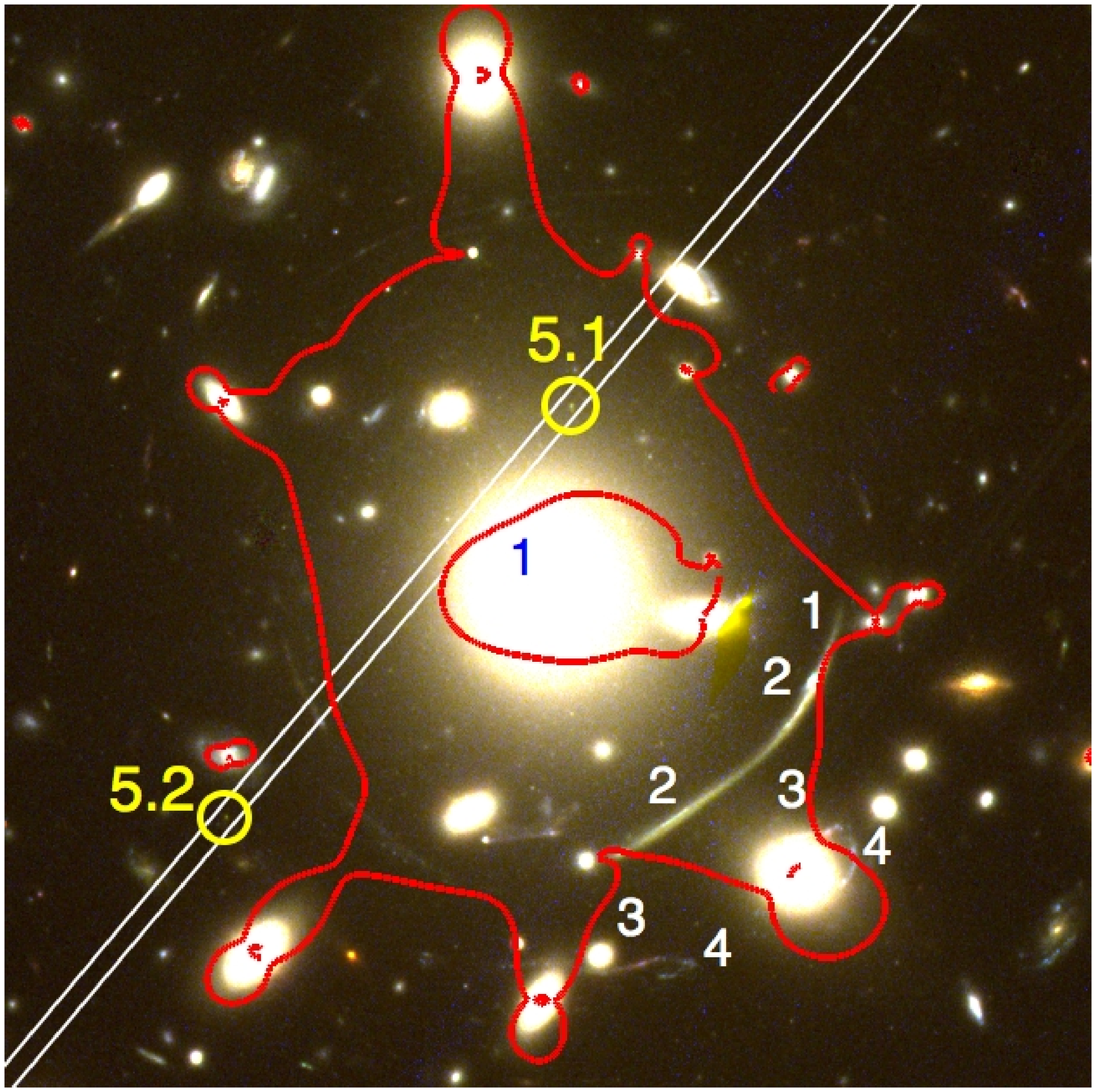}
\end{minipage}
\hspace{0.5cm}
\begin{minipage}{1cm}
\vspace{0.5cm}
5.1

\vspace{3.5cm}
5.2

\end{minipage}
\begin{minipage}{7.5cm}
\includegraphics[width=7.5cm,angle=0]{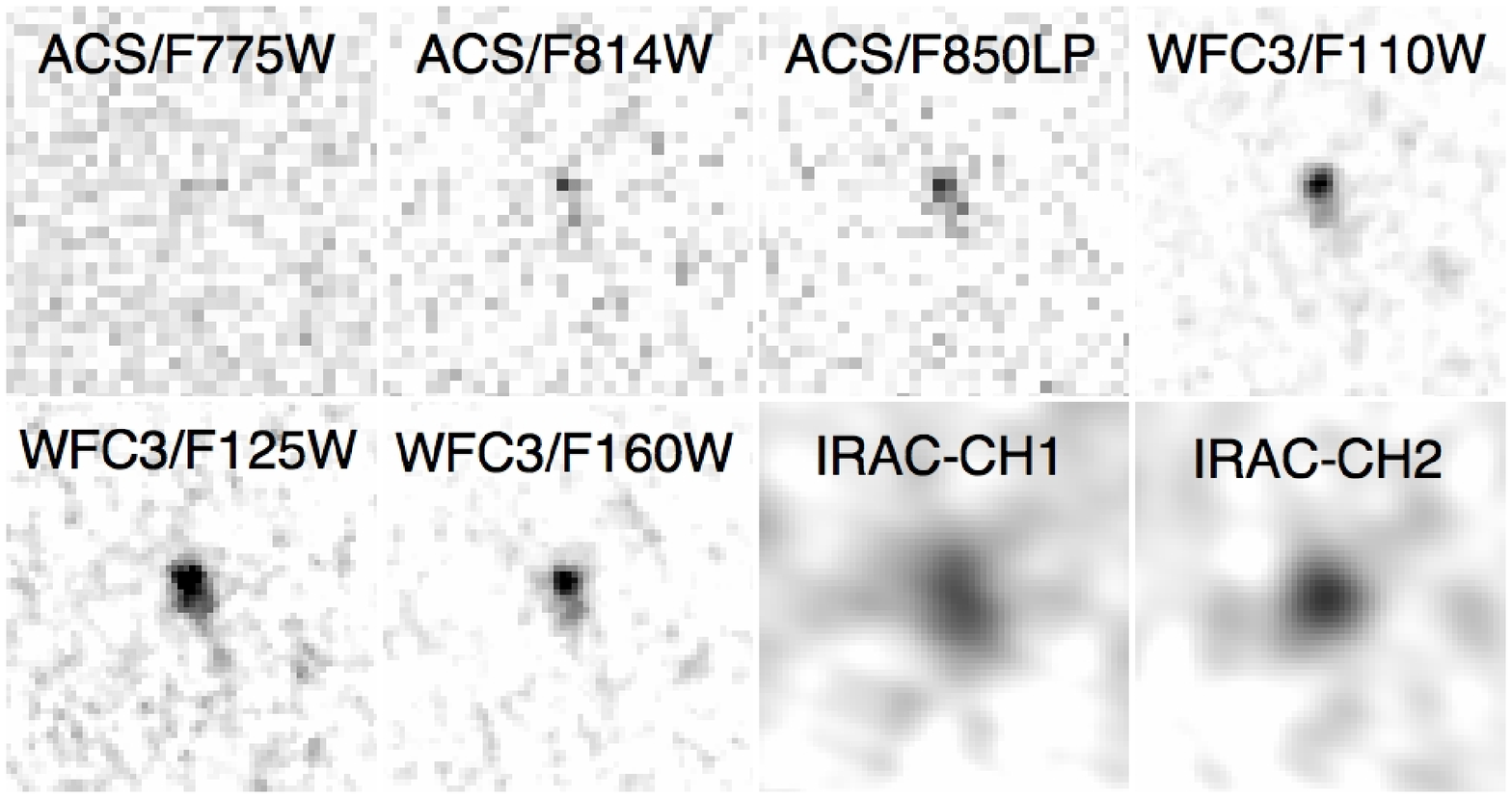}
\includegraphics[width=7.5cm,angle=0]{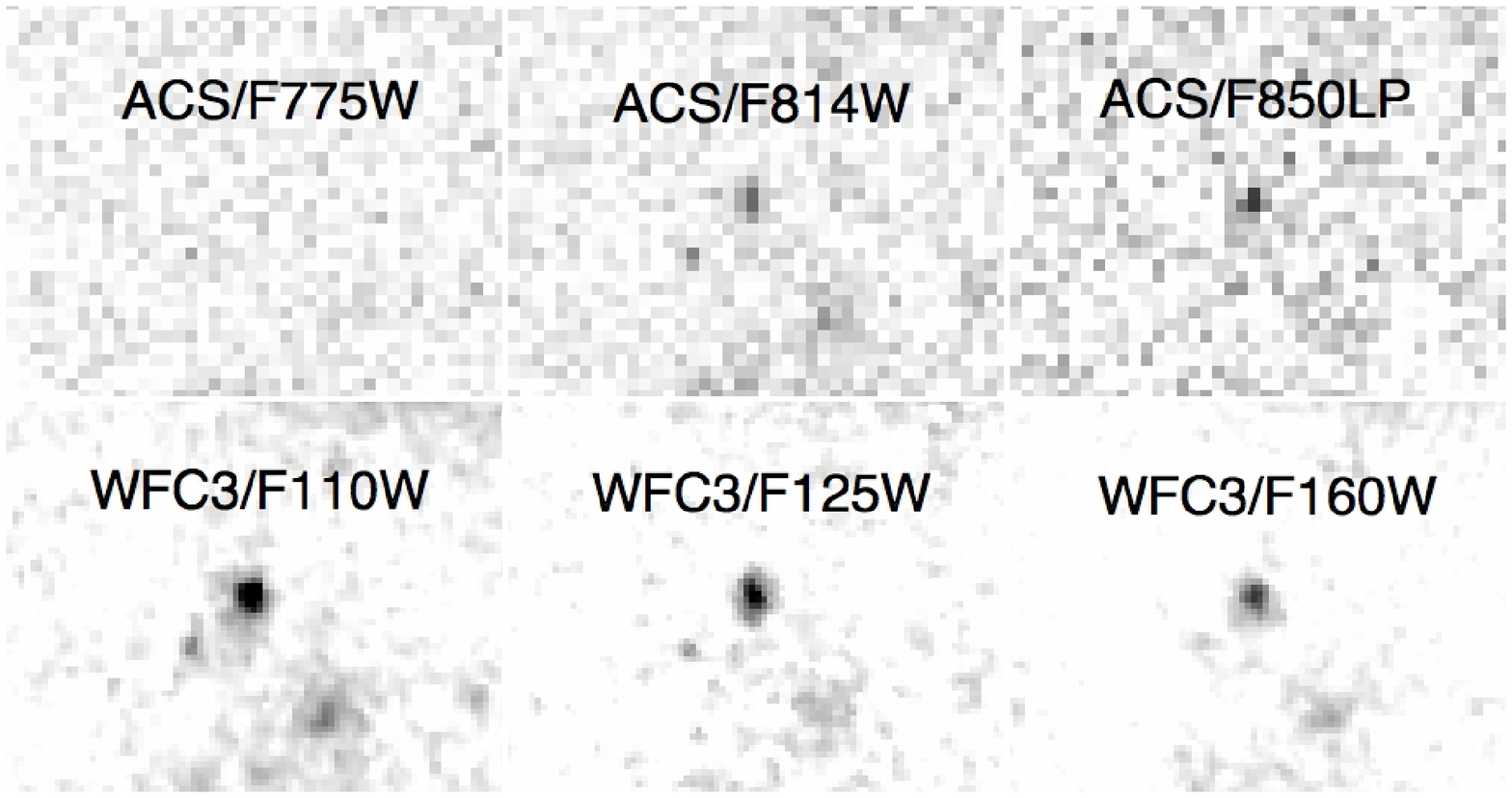}
\end{minipage}

\caption{\label{field}Left: F814W-F110W-F160W colour image of the
  very core of Abell 383, together with known multiply-imaged systems (marked 1 to 5) and 
the critical line at $z=6$. The identified i-dropout is seen as two
  images marked by yellow circles; the DEIMOS long-slit used 
  in our spectroscopic follow-up is shown in white. Right: postage stamps of both images. 
HST and IRAC images are 3\arcsec and 6\arcsec a side, respectively.
  }

\end{figure*}

HST images of the core of the massive
galaxy cluster Abell 383
($\alpha$=02:48:03.332, $\delta$=-03:31:44.98,$z=$0.187) have been
obtained with ACS and WFC3/IR between November 2010 and February 2011 in 10 filters from
optical to near-infrared (F435W, F475W, F606W, F625W, F775W, F814W, F850LP, F110W, F125W, 
F160W) with exposure times of 2000-3000\,s per band. Individual ACS frames
were corrected for charge transfer efficiency following the procedure
described by \citet{Massey-cte}, adjusted to correct the most recent
ACS images (more details will be given in Paraficz et al. in
prep.). We measured the relative shifts between different epochs of
observations in a given filter and combined the ACS and WFC3 with
multidrizzle \citep{Koekemoer} to obtain the final images.


We then performed a search for $i$ dropouts meeting the colour criteria F775W-F850LP$>1.3$ 
\citep{Bouwens06} in the strong-lensing region of the cluster. The envelope of the Brightest Cluster 
Galaxy (BCG) has been subtracted beforehand, using an elliptical surface-brightness model 
constructed with the {\sc IRAF} task {\sc ellipse} following the technique presented in \citet{Richard08}.
This model provides a good subtraction overall, except for residuals at the very centre 
that are due to the presence of a compact source and a dust lane, features already pointed 
out by \citet{Smith05}. We used  {\sc SExtractor} \citep{sextractor} in \textit{double-image} mode 
to obtain photometry across all HST bands. The F110W (most sensitive observation)  was 
used as a \textit{detection} image and the other bands as \textit{measurement} images.
Colours were measured between two bands inside the same 0.3\arcsec\ aperture and scaled 
to match the total magnitude in the F110W band (MAG\_AUTO with an isophote of $\sim0.6\arcsec$ 
diameter). We used a 3-$\sigma$ upper limit for the photometry 
in case of non-detections.

We identified two $i$-dropouts located 10\arcsec\ north (image 5.1:
$\alpha$=02:48:03.264, $\delta$=-03:31:34.77) and 25\arcsec\ south-east
(image 5.2: $\alpha$=02:48:04.600,$\delta$=-03:31:58.47) of the
BCG (Fig. \ref{field}). They show a strong continuum break (F775W-F850LP $>1.9$ mag.) 
and their colours agree within each other to better than the 1$\sigma$ error. The background gradient 
around image 5.1 is efficiently removed and the increased sky noise is estimated locally by {\sc SExtractor}. The measurements 
of both images are listed in Table \ref{photometry}. 
Our well constrained lensing model for the mass distribution in this cluster (see Section \ref{analysis}) 
predicts that both images 5.1 and 5.2 are associated with the same source (referred to as source 5), 
when assuming a redshift $z>5.5$. 

On January 3, 2011 we used the DEep Imaging Multi-Object Spectrograph
(DEIMOS) on the Keck II telescope to obtain an optical spectrum with a
1.2\arcsec-wide slit aligned to cover both images
(Fig. \ref{field}). Aiming to detect Lyman-$\alpha$ emission
from this source, we used the red-efficient 600ZD grating for three
exposures of 1800\,s. The reduction of the resulting data was performed with
standard {\sc IRAF} routines for bias subtraction, flat-fielding, sky
subtraction and wavelength calibration. The standard star Feige 110
observed during the same night was used for flux calibration. Seeing
was 1.1\arcsec and conditions photometric. 

\begin{figure}
\includegraphics[height=5.0cm,angle=0]{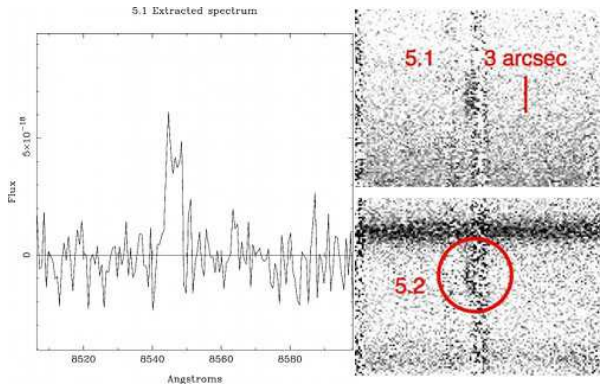}

\caption{\label{spectrum}2D long-slit spectrum as obtained with Keck/DEIMOS, covering both
  images 5.1 and 5.2 and showing a clear emission line at 8545
  \AA\ for both 5.1 and 5.2.  The inset presents the extracted spectrum of
  image 5.1 around this wavelength. }

\end{figure}

A strong
emission line is detected at 8545 \AA\ at the locations of both images in the
sky-subtracted spectrum (Fig. \ref{spectrum}). We measure an
integrated flux of $2.2\pm0.3\ 10^{-17}$ erg s$^{-1}$ cm$^{-2}$ in the extracted
1D spectrum of the brightest image (5.1) and a spatial extent of
$\sim$3\arcsec, much larger than the seeing of the observations. The
line profile shows a possible asymmetry, but due to the close proximity of a
skyline the signal-to-noise is too poor for an accurate measurement of the profile
on the red side. However, all available evidence (including the lensing
configuration and photometry) support the identification of this line as
 Lyman-$\alpha$ at a redshift $z_{\rm spec}=6.027\pm0.002$ (measured
at the peak of the line).



In addition to the HST images, we used the SST/IRAC data
of Abell 383 obtained at three different epochs (August 2005, September
2009, March 2010).  The first data set was obtained by programme 83
(PI: G.~Rieke) for a total integration time of 40 minutes/band.  The
second and third data sets were obtaind by programme 60034 (PI:
E.~Egami; ``The IRAC Lensing Survey'') and yielded a total 2-epoch integration
time of 5 hours/band.  The Basic Calibrated Data (BCD) frames were
mosaicked using {\sc MOPEX} (MOsaicker and Point source EXtractor), a software
package distributed by the \textit{Spitzer Science Center}, using a pixel scale
of 0\farcs6 (half of the original instrumental pixel scale).

Images 5.1 and 5.2 are detected in the two first IRAC channels
(3.6$\mu$m and 4.5$\mu$m). We used a model of the BCG convolved with the
IRAC point-spread function (PSF) to subtract its contribution and used
a larger aperture (2\arcsec) to measure the colour between the
WFC3/F110W band and the IRAC bands. Using a 1\arcsec aperture (more similar 
to the isophote used to infer the total magnitude) gives consistent colours 
but higher error bars. The WFC3 and IRAC images were convolved with an
empirical IRAC PSF and a {\sc tinytim} 
\citep{tinytim} model of the WFC3 PSF, respectively. 
We checked this photometry matching between HST and SST using cluster members as well 
as background sources of known spectroscopic redshift.
Again, the photometry for image 5.1
(scaled to the total magnitude in F110W) is listed in
Table~\ref{photometry}. IRAC colours were not measured for image 5.2 since we were 
unable to remove the contamination from nearby non-elliptical cluster members.  
As the observed IRAC colours are entirely
consistent between the 3 epochs, the BCG contribution to the sky noise is within the measured 
errors. We adopt an average of the values from the two best epochs (2 and 3). 

\section{Physical properties}

\label{analysis}


We now estimate the strong-lensing magnification factors of images 5.1 and 5.2 
based on a parametric model for the mass
distribution of the cluster core. This model is constructed using
{\sc lenstool}\footnote{\rm http://lamwws.oamp.fr/lenstool/} \citep{Jullo},
and based on a similar model constrained by 4 multiply imaged systems
(referred to as systems 1 to 4, Fig.~\ref{field}) presented previously by \citet{Smith03} and
\citet{Richard10b}.
Pseudo-isothermal elliptical mass distributions are used for both the
cluster-scale component and individual galaxies. We include in our 
optimisation the recent spectroscopic redshifts obtained by \citet{Newman11} 
for systems 3 and 4. The location of the new pair of images identified
in this letter (5.1 / 5.2) is in excellent agreement (within 0.5\arcsec) with the predictions from this
model at $z=6$. We therefore use this new information as an
additional constraint (system 5) to improve the mass model. From the family of
best models resulting from the optimisation, we derive a magnification
factor $\mu_1=11.4\pm1.9$ and $\mu_2= 7.3\pm1.2$ for images 5.1 and
5.2, respectively. The ratio $\mu_1/\mu_2$ is in perfect agreement with
the F110W photometry. A third image (5.3) is predicted but is
demagnified and located at the very centre of the BCG, making its
detection extremely challenging, if not impossible.

Taking into account these magnification factors, we derive intrinsic
properties for the original background source 5. Its unlensed F110W (J band) magnitude
(27.4 AB) corresponds to an $\sim$0.4$L^*$ galaxy at $z=6$, which
makes it more representative of the galaxy population at this redshift than brighter
objects, such as the i-dropout identified in Abell 1703
\citep{Richard09} which corresponds to a $\sim$3.6$L^*$ galaxy.
A particularly interesting feature of source 5 is the large magnitude break ($1.50\pm0.10$ mag)
between the H band (F160W) and the two first channels of IRAC, passbands
that straddle the location of the rest-frame Balmer break.
In order to interpret the implications of this break for the properties of
the underlying stellar populations, we performed a Spectral Energy
Distribution (SED) fit of image 5.1 (highest S/N image) with a library
of spectral templates following the precepts discussed in detail by
\citet{Stark09}. The
main parameters we aim to derive are the total stellar mass (M$^*$),
the current star-formation rate (SFR) and the age of the stellar
population ($T$). To this end, we removed the contribution of the
measured Lyman-$\alpha$ line in the F814W and F850LP photometry,
and then fit Charlot \& Bruzual(2007, S. Charlot, private
communication) stellar-population synthesis models to the observed
SEDs.  We considered exponentially decaying star formation histories
of the form $\rm{SFR(t)}\simeq exp(-t/\tau)$ with e-folding times of
$\tau=10$, 70, 100, 300, and 500 Myr, in addition to models with
continuous star formation (CSF). $T$ is restricted to the range of 10 to 1000 Myr 
(the age of the universe at $z\sim6$).  We use a Salpeter 
initial mass function (IMF) \citep{Salpeter} and the dust extinction
law of \citet{Calzetti} with $0.0<E(B-V)<1.0$. Finally, we allow the metallicity to vary between solar
(Z$_\odot$) and 0.2 Z$_\odot$, but note that metallicity is degenerate with
reddening and has a second-order effect on the SED.  Absorption by the 
intergalactic medium is accounted for following \citet{Meiksin}.


We next consider the possibility that the strong break between the H band 
and the IRAC channels is due
to contamination by nebular emission lines. Strong H$\alpha$, H$\beta$
and $[OIII]$ lines would all fall into the IRAC bands and could boost the
continuum, mimicking a stronger Balmer break \citep{Schaerer10}. We
correct for this effect as follows: the SFR estimated from
SED fitting is consistent with the observed Lyman-$\alpha$ flux, when
assuming an aperture correction of a factor of 2 and (conservatively)
1/3 case B for Lyman-$\alpha$. We predict observed fluxes
$f_{H\alpha}= 1.46$ $10^{-17}$, $f_{H\beta}=5.1$ $10^{-18}$ and
$f_{\rm[OIII]}=6.7$ $10^{-18}$ (cgs units, both [OIII] lines average
from COSMOS galaxies, Zoubian et al. 2011 in prep.). We therefore
 lower the fluxes by 0.06 and 0.04 magnitudes at
3.6 $\mu$m and 4.5$\mu$m, respectively, thereby eliminating contamination
from unresolved line emission. This correction is negligible and we observe no significant
change in the best-fit model using this updated photometry.
 
\begin{figure}
\includegraphics[width=8cm,angle=0]{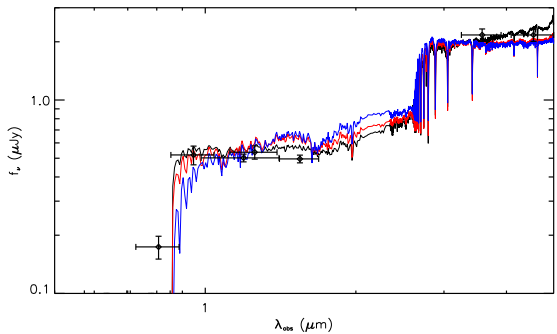}

\caption{\label{sed}Observed SED of image
  5.1 (black data points) and model templates. The black curve 
  represents the best fit to the
  photometry ($\tau=500$ Myrs), whereas the blue/red curves show the best fit 
  for younger but more punctuated star-formation histories ($\tau=10/100$ Myrs respectively), giving 
  a poorer fit to the photometry.}
\end{figure}

 The observed SED and the best-fitting template are presented in
 Fig.~\ref{sed}. The best fit is achieved for a $\tau=500$ Myr
 star-formation history. The full set of templates gives an age range
 of $T$=640-940 Myrs (within a $\Delta\chi^2=3$ from the best model), 
 depending on the exact star-formation history, corresponding to a redshift of formation of 
 $z_f=18\pm4$. More punctuated star-formation histories ($\tau=10-100$ Myrs) could produce 
 similar Balmer breaks with ages of only $T=130-300$ Myrs, but they produce much 
 poorer fits to the data ($\Delta\chi^2\sim50$, Fig.~\ref{sed}). However, we acknowledge these 
 are simple models and age-dependent dust extinction \citep{Poggianti} as well as 
more complex star-formation histories (two-component models, 
 with bursts on top of a continuous star-formation) could provide younger ages. This would 
affect the inferred ages more strongly than the inferred stellar masses, which are more independent 
of the assumed star-formation history. The most precise measurement is the rest-frame V-band  
luminosity of 2.2$^{+0.97}_{-0.42}$ $10^{10}$ L$_\odot$. The inferred values from the best fit model are a stellar
 mass of M$^*$=6.3$^{+2.8}_{-1.2}$ $10^{9}$ M$_\odot$, current SFR
 of 3.2$\pm$1.1 M$_\odot$/yr, and specific SFR ($SSFR= SFR/M^*$,
 e.g. \citealt{Brinchmann2004}) of $\sim5$ $10^{-10}$ yr$^{-1}$
 indicate relatively vigorous star formation similar to the one found
 by \citet{Egami}. Note that the relative error in the magnification factor has
 been added in quadrature for all intrinsic values.  The rest-frame
 UV photometry, around 1500 \AA, is consistent with a model of
 constant AB magnitude, or equivalently with a slope of $\beta\sim2$ (defined
 as $f_\lambda\propto\lambda^{-\beta}$). This is consistent with the
  very small amount of dust (E(B-V)$<0.06$) in the best-fit model, and also 
  with the results found
 for $z\sim7$ dropouts in the UDF \citep{Bouwens_slope}. 

Interestingly, image 5.1 of the galaxy is well resolved in the
WFC3/F110W image, as an ellipse of size
$\sim$0.55\arcsec$\times$0.3\arcsec based on well detected
pixels.  We measure a half-light radius $r_h=0.21\arcsec$ with
{\sc SExtractor}, corresponding to a source-plane radius of 0.06\arcsec, or
0.3 kpc after correcting for the PSF contribution. In comparison, the 3\arcsec
extent of Lyman-$\alpha$ emission observed with DEIMOS, at least in
the direction of the slit orientation, corresponds to a 5 kpc diameter
in the source plane. This discrepancy consititutes strong evidence for the presence of a
large envelope of neutral hydrogen surrounding the star-forming region
responsible for the UV emission \citep{Steidel2011}. Of course, as these size measurements 
are given in UV rest-frame, it is quite likely that the source is imbedded in a lower surface-
brightness halo, and the low reddening inferred would be coming from the UV-bright component. Nevertheless, 
a possible 3\arcsec\ diameter dusty component would have been resolved with IRAC.

\section{An evolved galaxy in a young Universe ?}

Taken at face value, the presence of a large optical break and an inferred mature stellar population
($\sim$800 Myrs) only 1 Gyr after the Big Bang is challenging. As previously discussed, this 
value depends on the assumptions made in the standard models of star-formation history used.
\citet{Mobasher} reported a similarly strong Balmer break for a brighter and more
massive galaxy at $z\sim6.5$, which, however, still lacks spectroscopic confirmation
despite numerous attempts. If we compare the intrinsic (i.e., demagnified) properties of this
lensed galaxy with those of the 16 $i$-dropouts with spectroscopic
redshifts analysed by \citet{Eyles} and fit with similar SED models, this galaxy is 1 magnitude
fainter in rest-frame UV ($F850LP=27.4$) but has a similar flux
redward of the Balmer break (although with a much stronger detection thanks
to its magnification by the cluster lens) and therefore the same typical mass.  The small
size of the star-forming region implies a star-formation surface
density of $\sim$10 M$_{\odot}$ yr$^{-1}$ kpc$^{-2}$, comparable to the most 
intense starburst activity observed locally \citep{Kennicutt} and allowing this object 
to drive strong winds.

The detection of a large Balmer break, at such a high significance,
is quite remarkable and demonstrates that sub-luminous systems (when
selected in the UV) may not always be observed at a young stage of
evolution. This is consistent with results obtained from IRAC stacking of faint $z=6-7$ field
galaxies \citep{Stark09,Labbe10}, and 15\% of the \citet{Stark09} sample of bright 
$i$-dropouts show similar Balmer breaks. Thanks to the magnification provided
by gravitational lensing, we can measure this feature in faint individual
objects. This result is important both for estimating the mass
function at high redshift, and even more for understanding the process
of reionisation, as it implies a significant contribution at $z>10$
from the progenitors of these objects.

From our knowledge of the UV luminosity function at $z\sim6$
\citep{Bouwens06}, we expect $\sim0.5$ objects of this kind per CLASH cluster
in the strong-lensing (multiply imaged) regime down to 0.2
L$^*$. Deeper WFC3 and ACS observations of strong-lensing clusters
would efficiently make use of the \textit{Spitzer Warm mission}
program and allow us to measure the distribution of Balmer breaks in a
larger sample down to even fainter magnitudes, and thus to precisely determine
the mass function.

\section*{Acknowledgments}
We thank the referee for a report that improved the presentation of our results. 
JR acknowledges support from the Dark Cosmology Centre. JPK
acknowledges support from the CNRS. DPS acknowledges support from
STFC. The Dark Cosmology Centre is funded by the Danish National
Research Foundation. Data presented herein were obtained as part of a
Multi-Cycle Program from the NASA/ESA \textit{Hubble Space Telescope}
(\#12065).  This work is based in part on observations made with the
\textit{Spitzer Space Telescope}, which is operated by the Jet Propulsion
Laboratory, California Institute of Technology under a contract with
NASA, and support was provided by NASA through an award issued by
JPL/Caltech.  Part of the work is also based on observations at the
W.M. Keck Observatory. The authors recognise and acknowledge the very
significant cultural role and reverence that the summit of Mauna Kea
has always had within the indigenous Hawaiian community. We are most
fortunate to have the opportunity to conduct observations from this
mountain. This work received support from Agence Nationale de la Recherche 
bearing the reference ANR-09-BLAN-0234-01. 

\bibliography{references}

\label{lastpage}

\end{document}